\documentclass[sigconf,screen]{acmart}

\usepackage{amsmath,amsfonts}
\usepackage{algorithmicx}
\usepackage{textcomp}
\usepackage{url}
\usepackage{listings, multicol}
\usepackage{upquote}
\usepackage{xspace}
\usepackage[utf8]{inputenc}
\usepackage[T1]{fontenc}
\usepackage{microtype}
\usepackage[english]{babel}
\usepackage{amsthm}
\usepackage{multirow}
\usepackage[para,online,flushleft]{threeparttable}
\usepackage[framemethod=TikZ]{mdframed}
\usepackage{subfigure}
\usepackage{balance}
\usepackage{enumitem}

\mdfdefinestyle{MyFrame}{%
    linecolor=black,
    outerlinewidth=0.3pt,
    roundcorner=5pt,
    skipabove = 6.5pt,
    innertopmargin=6.5pt, 
    innerbottommargin=6.5pt, 
    innerrightmargin=6.5pt,
    innerleftmargin=6.5pt,
    backgroundcolor=gray!25!white
}

\definecolor{lightgray}{rgb}{0.95, 0.95, 0.95}
\definecolor{darkgray}{rgb}{0.4, 0.4, 0.4}
\definecolor{purple}{rgb}{0.65, 0.12, 0.82}
\definecolor{editorGray}{rgb}{0.95, 0.95, 0.95}
\definecolor{editorOcher}{rgb}{1, 0.5, 0} 
\definecolor{editorGreen}{rgb}{0, 0.5, 0} 
\definecolor{orange}{rgb}{1,0.45,0.13}
\definecolor{olive}{rgb}{0.17,0.59,0.20}
\definecolor{brown}{rgb}{0.69,0.31,0.31}
\definecolor{purple}{rgb}{0.38,0.18,0.81}
\definecolor{lightblue}{rgb}{0.1,0.57,0.7}
\definecolor{lightred}{rgb}{1,0.4,0.5}
\definecolor{codegreen}{rgb}{0,0.6,0}
\definecolor{codegray}{rgb}{0.5,0.5,0.5}
\definecolor{codepurple}{rgb}{0.58,0,0.82}
\definecolor{backcolour}{rgb}{0.95,0.95,0.92}

\lstdefinelanguage{JavaScript}{
    keywords={typeof, new, true, false, catch, function, return, null, catch, switch, var, if, in, while, do, else, case, break},
    keywordstyle=\color{blue}\bfseries,
    ndkeywords={class, export, boolean, throw, implements, import, this},
    ndkeywordstyle=\color{darkgray}\bfseries,
    identifierstyle=\color{black},
    sensitive=false,
    comment=[l]{//},
    morecomment=[s]{/*}{*/},
    commentstyle=\color{purple}\ttfamily,
    stringstyle=\color{red}\ttfamily,
    morestring=[b]',
    morestring=[b]"
}

\lstdefinestyle{codestyle}{
    backgroundcolor=\color{white},
    commentstyle=\color{codegreen},
    keywordstyle=\color{magenta},
    numberstyle=\tiny\color{codegray},
    stringstyle=\color{codepurple},
    basicstyle=\ttfamily\small,
    breakatwhitespace=false,
    breaklines=true,
    captionpos=b,
    keepspaces=true,
    numbers=left,
    numbersep=5pt,
    showspaces=false,
    showstringspaces=false,
    showtabs=false,
    tabsize=2,
    lineskip=-.1cm,
    xleftmargin=0.4cm
}

\newcommand{\todoorange}[1]{\todoc{orange!60}  {[#1]}}
\newcommand{\todocyan}[1]{\todoc{cyan}  {[#1]}}
\newcommand{\todoviolet}[1]{\todoc{violet}  {[#1]}}

\newcommand{\todoc}[2]{}
\newcommand{\aaron}[1]{\todoorange{Aaron: #1}}

\newcommand{\scc}[1]{\todoviolet{scc: #1}}
\newcommand{\lili}[1]{\todocyan{Lili: #1}}

\newcommand{\addition}[1]{#1}
\newcommand{\deletion}[1]{}

\theoremstyle{non-determinism}
\newtheorem{non-determinism}{Non-determinism}
\theoremstyle{definition}

\theoremstyle{non-determinism}

\theoremstyle{non-determinism}
\newtheorem{assertion}{Assertion}
\theoremstyle{non-determinism}

\makeatletter
\def\namedlabel#1#2{\begingroup
    #2%
    \def\@currentlabel{#2}%
    \phantomsection\label{#1}\endgroup
}

\newcommand{\CreatedState}{\textit{Created}\xspace}
\newcommand{\PendingState}{\textit{Pending}\xspace}
\newcommand{\ExecutedState}{\textit{Executed}\xspace}
\newcommand{\ReversedState}{\textit{Reversed}\xspace}
\newcommand{\ConfirmedState}{\textit{Finalized}\xspace}
\newcommand{\DroppedState}{\textit{Dropped}\xspace}

\newcommand{\darcher}{\textit{\DJ Archer}\xspace}
\newcommand{\bug}{on-chain-off-chain synchronization bug\xspace}
\newcommand{\bugs}{on-chain-off-chain synchronization bugs\xspace}

\newcommand{\Bugs}{On-chain-off-chain synchronization bugs\xspace}
\newcommand{\DApp}{DApp\xspace}
\newcommand{\DApps}{DApps\xspace}

\newcommand{\TPS}{16.4\xspace}
\newcommand{\txSubmitRate}{32.5\xspace}

\newcommand{\uncleHeightInterval}{24.43\xspace}
\newcommand{\uncleTimeInterval}{11.06\xspace}

\newcommand{\numRevokedTransactionsPerHour}{16.69\xspace} 

\newcommand{\numSubjects}{11\xspace}

\newcommand{\numDetectedBugs}{15\xspace} 
\newcommand{\precision}{99.3\%\xspace}
\newcommand{\recall}{87.6\%\xspace}
\newcommand{\accuracy}{89.4\%\xspace}
\newcommand{\numReportedBugs}{15\xspace}
\newcommand{\numConfirmedBugs}{six\xspace} 
\newcommand{\numFixedBugs}{three\xspace}

\makeatletter

\newcommand{\Rmnum}[1]{\expandafter\@slowromancap\romannumeral #1@}
\makeatother

\AtBeginDocument{%
    \providecommand\BibTeX{{%
        \normalfont B\kern-0.5em{\scshape i\kern-0.25em b}\kern-0.8em\TeX}}}

\copyrightyear{2021} 
\acmYear{2021} 
\setcopyright{none}
\acmConference[ESEC/FSE '21]{Proceedings of the 29th ACM Joint European Software Engineering Conference and Symposium on the Foundations of Software Engineering}{August 23--28, 2021}{Athens, Greece}
\acmBooktitle{Proceedings of the 29th ACM Joint European Software Engineering Conference and Symposium on the Foundations of Software Engineering (ESEC/FSE '21), August 23--28, 2021, Athens, Greece}
\acmPrice{15.00}
\acmDOI{10.1145/3468264.3468546}
\acmISBN{978-1-4503-8562-6/21/08}
\acmSubmissionID{fse21main-p98-p}

\begin{document}

    \title{\darcher: Detecting On-Chain-Off-Chain Synchronization Bugs in Decentralized Applications}

\author{Wuqi Zhang}
\orcid{0000-0001-8039-0528}
\affiliation{%
 \institution{Department of Computer Science and Engineering, The Hong Kong University of Science and Technology}
 \city{Hong Kong}
 \country{China}
}
\email{wzhangcb@cse.ust.hk}
\author{Lili Wei}
\orcid{0000-0002-2428-4111}
\authornote{Lili Wei is the corresponding author of this paper.}
\affiliation{%
 \institution{Department of Computer Science and Engineering, The Hong Kong University of Science and Technology}
 \city{Hong Kong}
 \country{China}
}
\email{liliwei@cse.ust.hk}
\author{Shuqing Li}
\orcid{0000-0001-6323-1402}
\affiliation{%
 \institution{Department of Computer Science and Engineering, Southern University of Science and Technology}
 \city{Shenzhen}
 \state{Guangdong}
 \country{China}
}
\email{lisq2017@mail.sustech.edu.cn}

\author{Yepang Liu}
\orcid{0000-0001-8147-8126}
\affiliation{%
 \institution{Department of Computer Science and Engineering, Guangdong Provincial Key Laboratory of Brain-inspired Intelligent Computation,
 Southern University of Science and Technology,}
 \city{Shenzhen}
 \state{Guangdong}
 \country{China}
}
\email{liuyp1@sustech.edu.cn}

\author{Shing-Chi Cheung}
\orcid{0000-0002-3508-7172}
\affiliation{%
 \institution{Department of Computer Science and Engineering, The Hong Kong University of Science and Technology}
 \city{Hong Kong}
 \country{China}
}
\email{scc@cse.ust.hk}

%
%
%
%


    \begin{abstract}
       Since the emergence of Ethereum, blockchain-based decentralized applications (DApps) have become increasingly popular and important.
       To balance the security, performance, and costs, a DApp typically consists of two layers: an on-chain layer to execute transactions and store crucial data on the blockchain and an off-chain layer to interact with users.
       A DApp needs to synchronize its off-chain layer with the on-chain layer proactively. 
       Otherwise, the inconsistent data in the off-chain layer could mislead users and cause undesirable consequences, e.g., loss of transaction fees.
       However, transactions sent to the blockchain are not guaranteed to be executed and could even be reversed after execution due to chain reorganization.
       Such non-determinism in the transaction execution is unique to blockchain. 
       DApp developers may fail to perform the on-chain-off-chain synchronization accurately due to their lack of familiarity with the complex transaction lifecycle.

       In this work, we investigate the challenges of synchronizing on-chain and off-chain data in Ethereum-based DApps.
       We present two types of bugs that could result in inconsistencies between the on-chain and off-chain layers.
       To help detect such on-chain-off-chain synchronization bugs, we introduce a state transition model to guide the testing of DApps and propose two effective oracles to facilitate the automatic identification of bugs.
       We build the first testing framework, \darcher, to detect on-chain-off-chain synchronization bugs in DApps.
       We have evaluated \darcher on \numSubjects popular real-world DApps.
       \darcher achieves high precision (\precision), recall (\recall), and accuracy (\accuracy) in bug detection and significantly outperforms the baseline methods.
       It has found \numDetectedBugs real bugs in the 11 DApps.
       So far, \numConfirmedBugs of the \numDetectedBugs bugs have been confirmed by the developers, and \numFixedBugs have been fixed. These promising results demonstrate the usefulness of \darcher.
    \end{abstract}



    \maketitle

    \section{Introduction}
\label{sec:introduction}


\underline{D}ecentralized \underline{App}lications (DApps) are software applications running on a decentralized network like blockchain.
Since the emergence of Ethereum~\cite{ethereum}, a blockchain platform that supports Turing-complete smart contracts, blockchain-based DApps have drawn much attention from both academia and industry.
\addition{As of April 2021, on Ethereum, there are over 2.7 thousand DApps and 70 thousand active DApp users, issuing over 170 thousand transactions per day~\cite{stateofthedappsStateDAppsDApp2021}.}

\begin{figure}
    \centering
    \includegraphics[scale=0.5]{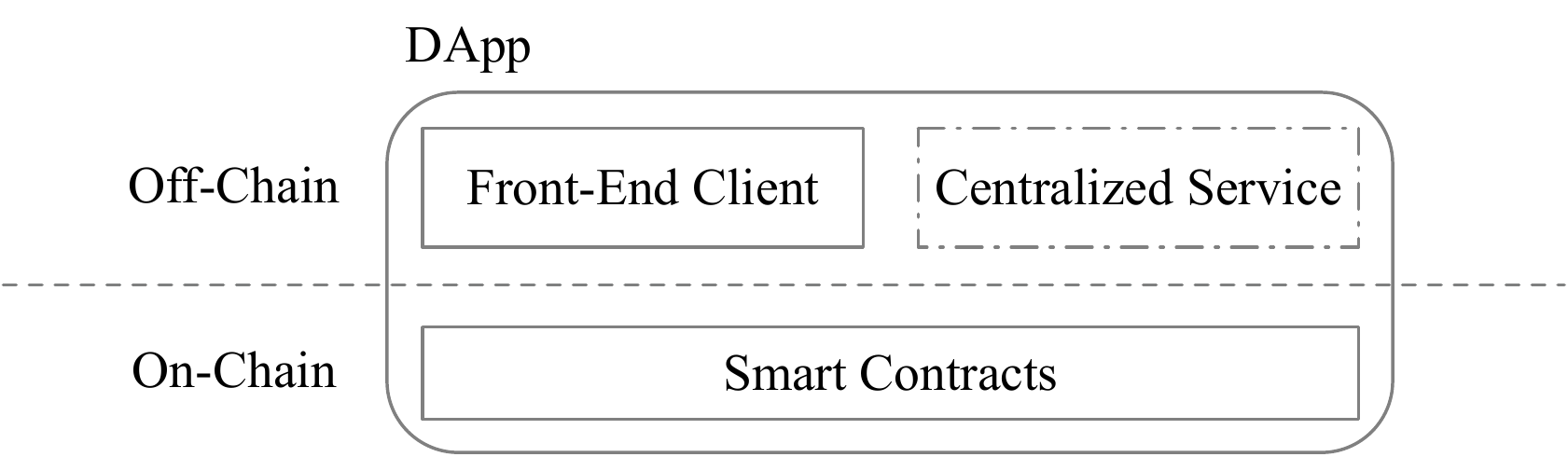}
    \caption{The Architecture of A Typical DApp}
    \label{fig:dapp-architecture}
\end{figure}

As shown in Fig.~\ref{fig:dapp-architecture}, DApps are typically deployed as web applications, consisting of two layers: \textit{on-chain} and \textit{off-chain}.
The former comprises a set of smart contracts that store and update crucial data on the blockchain.\footnote{To save transaction costs, blockchain typically only stores crucial data~\cite{wuFirstLookBlockchainbased2019}.}
The latter contains a user-friendly front-end client and an optional centralized service outside the blockchain~\cite{wesslingBlockchainTacticsBuilding2019}.
Take Giveth~\cite{giveth}, a popular DApp for charitable donation on Ethereum, as an example.
It comprises a set of on-chain smart contracts and an off-chain front end backed by a centralized cache server.
To avoid frequent communication with the blockchain and facilitate users' queries, the off-chain layer usually stores processed or analyzed results of certain important on-chain data.
Therefore, a DApp's state is composed of an \textit{on-chain state} and an \textit{off-chain state}, referring to the data stored at the on-chain (e.g.,\ data in Giveth's smart contracts) and the off-chain  (e.g.,\ database of Giveth's cache server) layers, respectively.
The on-chain and off-chain states of a DApp are not necessarily the same.
DApps usually maintain a mapping between the on-chain and off-chain states.
Since blockchain is autonomous and the on-chain state may change out of the control of DApps, DApps need to proactively synchronize their off-chain states and keep them consistent with the on-chain states. 
We call such a process \textit{on-chain-off-chain synchronization}.

On-chain-off-chain synchronization can be complicated.
Changes are made to the on-chain state of a DApp by sending transactions to the blockchain.
However, transaction executions are non-deterministic due to the decentralized nature of blockchain~\cite{zhangAnalysisMainConsensus2020}.
On the one hand, transactions sent to the blockchain are not necessarily executed and could be dropped silently\addition{ even after being acknowledged by the miners}~\cite{cancelTx}.
On the other hand, executed transactions could be reversed as a result of chain reorganization~\cite{liSurveySecurityBlockchain2020}.
Such non-determinism is unique to blockchains and does not exist in conventional centralized services or distributed systems.
If the non-determinism is not carefully considered and dealt with in the development of a DApp,
inconsistencies between its off-chain and on-chain states may arise when the DApp is running.
Such inconsistencies can cause catastrophic consequences.
It is because users typically take actions (e.g., buying and selling) according to the off-chain state shown at the front-end client of a DApp.
Stale data at the off-chain layer can mislead users into taking wrong actions that result in irreversible changes on the blockchain or financial losses.
For example, in a DApp like Giveth, if a donation transaction is reversed on the blockchain while the DApp's off-chain layer is unaware of the reverse and does not update the donation status, subsequent transactions to withdraw the donation will fail, causing the loss of transaction fees.\footnote{Ethereum users need to pay fees for each transaction, no matter the transaction succeeds or not~\cite{yellowpaper}.}
We refer to such bugs that stem from non-deterministic transaction execution and induce inconsistencies between the on-chain and off-chain states as \textit{on-chain-off-chain synchronization bugs}.

\addition{
It is non-trivial to avoid \bugs even if developers have considered them in the development of \DApps. 
Giveth developers have written more than 600 lines of code}\footnote{\url{https://github.com/Giveth/feathers-giveth/blob/d51d585/src/blockchain/watcher.js}}\addition{ to track the lifecycle of all transactions and keep the consistency between the on-chain and the off-chain states. 
Augur~\cite{augur}, another very popular \DApp, implements a delicate rollback table}\footnote{\url{https://github.com/AugurProject/augur/blob/85b570f/packages/augur-sdk/src/state/db/RollbackTable.ts}}\addition{ to revert off-chain states when the transactions are reversed \lili{reversed or reverted?}\aaron{We specially use ``reverse'' for transactions, as in the lifecycle model.} on the blockchain.
}
\addition{
In spite of the efforts that developers take, we still find \bugs in those \DApps, as discussed in Section~\ref{subsec:results-for-rq:usefulness}.
}

While several bug detection techniques have been proposed to assure the quality of DApps, they are either general tools for specifying test cases and assertions~\cite{wuKayaTestingFramework2020} or focus only on the on-chain layer by finding defects in smart contracts~\cite{luuMakingSmartContracts2016,jiangContractFuzzerFuzzingSmart2018,grechMadMaxSurvivingOutofgas2018,wangDetectingNondeterministicPayment2019,nguyenSFuzzEfficientAdaptive2020}.
In other words, none of them can effectively pinpoint \bugs in DApps.
This motivates us to investigate the \bugs and devise testing techniques to detect such bugs.

\Bugs could occur in DApps on all blockchain platforms.
Due to the impact of the Ethereum blockchain, our work focuses on Ethereum-based DApps.
To ease presentation, we may simply refer to Ethereum-based DApps as DApps in the following.

There are two major challenges in detecting \bugs in DApps.
First, we need to expose the non-determinism of the transaction executions, which is not considered by existing testing environments, so that on-chain-off-chain synchronization bugs can be better revealed.
Second, \addition{the mapping between on-chain and off-chain states is usually unavailable and hard to specify. 
Hence, it is impractical to compare on-chain and off-chain states to check consistency directly.} 
We need a decidable criterion to mechanically judge \addition{the consistency, thus identifying} the existence of the bugs. 
To address the first challenge, we study the causes of the synchronization bugs and propose a state transition model for the lifecycle of transactions on the Ethereum blockchain.
Guided by the model, we are able to trigger the scenarios where transactions are dropped or reversed to test DApps.
To address the second challenge, we propose two test oracles based on the following key observation: the off-chain state of a DApp should stay the same if the corresponding on-chain state is unchanged, and a violation of this requirement would indicate the existence of bugs in the state synchronization process.
With these oracles, we are able to define assertions only on the off-chain states to effectively reveal \bugs in DApps, \addition{without knowing the mapping between the on-chain and the off-chain states}.

We implement our approach as a DApp testing framework, called \darcher, and evaluate it using \numSubjects popular real-world DApps.
These DApps vary in scale, ranging from hundreds to tens of thousands of lines of code, and have received at least 100 stars on GitHub.
Our experiment results show that \darcher is effective. 
It is able to detect \bugs in all of the DApp subjects.
We manually check warnings reported by \darcher, group the ones with the same root cause, and submit \numReportedBugs issue reports on GitHub.
So far, bugs in \numConfirmedBugs issue reports have been confirmed by developers.
These bugs can cause transaction failures, continuous errors being prompted in the DApp client UI, or incorrect information being displayed to users.

This paper makes three major contributions:

\begin{itemize}[leftmargin=*, itemsep = 3pt, topsep = 3pt]
    \item To the best of our knowledge, this is the first study that examines the bugs in the on-chain-off-chain synchronization process of DApps.
    We formulate the problem with a state transition model of the transaction lifecycles.
    Using the model, we present the challenges in on-chain-off-chain synchronization and the causes of the synchronization bugs.

    \item We propose two test oracles \addition{that are able to effectively identify inconsistencies between on-chain and off-chain states without knowing the mapping between them.}
    Based on the transition model and the proposed oracles, a testing framework, \darcher, is built to detect on-chain-off-chain synchronization bugs in Ethereum-based \DApps.
    \darcher is open-source on GitHub\footnote{\url{https://github.com/Troublor/darcher}}.

    \item We evaluate \darcher on \numSubjects real-world DApps and conclude that \darcher can detect bugs with high precision (\precision), recall (\recall), and accuracy (\accuracy).
    We submit \numReportedBugs issue reports to developers, and \numConfirmedBugs have been confirmed.

\end{itemize}


    \section{Background}\label{sec:background}

\subsection{Two-Layer Architecture of DApps}\label{subsec:on-chain-and-off-chain-layers-of-dapps}
In Section~\ref{sec:introduction}, we have briefly introduced the architecture of DApps.
This section further explains several important concepts in detail.

\subsubsection{On-Chain and Off-Chain Layers}
The two-layer architecture of DApps is to balance security, maintainability, performance, and costs~\cite{wesslingBlockchainTacticsBuilding2019}.
Blockchain (e.g.,~Ethereum~\cite{yellowpaper}), as a decentralized ledger, offers a highly secure data store with programmable smart contracts in the on-chain layer.
However, storing data and computations on blockchain incurs a high latency and requires paying a non-negligible transaction fee.
Besides, interacting with blockchain \addition{by sending transactions and interpreting logs in low-level bytecode} is also not friendly for ordinary users.
The off-chain layer is meant to improve the performance and reduce costs using a user-friendly front-end and centralized services (optional) without sacrificing too much in the way of security~\cite{wesslingHowMuchBlockchain2018}.

\subsubsection{On-Chain and Off-Chain States}
As mentioned in Section~\ref{sec:introduction}, DApps execute key logics implemented in smart contracts and store crucial data on the blockchain as the on-chain state. 
Users usually take actions in the off-chain layer according to the results of some calculations involving on-chain data.
It is expensive to store such intermediate results back onto the blockchain.
It is also inefficient to query the on-chain state to perform the calculations repetitively.
As a result, DApps usually store a simplified view of their on-chain state and some relevant calculation results at the off-chain layer as the off-chain state.
Note that the off-chain layer may also store other data irrelevant to the on-chain state.
In this paper, we do not include such data in our definition of the off-chain state.

\subsubsection{On-Chain-Off-Chain Synchronization}
DApp users send transactions to the blockchain via the off-chain layer to make changes to the on-chain state.
During the lifecycle of each transaction, blockchain emits various events.
DApps can monitor such events to keep their off-chain state synchronized with the corresponding on-chain state.
This process is easy to implement if the transactions are executed deterministically on a centralized service or database.
However, this is not the case on the blockchain, as will be explained in the next subsection.


\subsection{Non-deterministic Transaction Execution}\label{subsec:transactions-on-ethereum}
Transactions sent to the blockchain will be broadcast to miners throughout the network.
Conceptually, miners on a blockchain collectively maintain a pool of transactions awaiting execution.
A transaction is added to the transaction pool when it is received by a miner.
A transaction is executed non-deterministically on the blockchain from two aspects.
First, a transaction may not be executed after it has been sent by DApp users.
Second, an executed transaction can be reversed from the transaction history.
In the following, we explain how such non-determinism arises. 

\subsubsection{Sent Transactions May Not Be Executed}
A miner can select which transactions to execute from the pool when a new block has been mined.
It is possible that a transaction is not selected and keeps staying in the pool.
An old transaction in the pool can be invalidated by a new one with the same nonce, which is the index of a transaction sent from the user~\cite{yellowpaper,cancelTx}.
The invalidated transactions will be dropped and never be executed on the blockchain.
In addition, miners can silently drop transactions for various reasons (e.g., due to the size limit of the transaction pool).

\subsubsection{Executed Transactions May Be Reversed}
A newly mined block is not necessarily added to the blockchain.
When multiple miners concurrently mine a new block~\cite{zhangAnalysisMainConsensus2020},
the blockchain will fork multiple chains of blocks. 
To resolve this problem, the blockchain will validate only the longest chain and invalidate the others. 
The process is known as \textit{chain reorganization}~\cite{nakamotoBitcoinPeertoPeerElectronic2008}.
If the execution of a transaction is recorded by a block in an invalidated chain, the transaction will be reversed and put back to the pool.
In practice, chain reorganization can hardly affect blocks with a sufficient number of \textit{confirmations}, which are the succeeding blocks on the same chain~\cite{gervaisSecurityPerformanceProof2016}.

Such non-determinism in transaction execution complicates the interactions and synchronization between the on-chain and off-chain layers of a DApp.
Bugs can arise if the DApp handles the non-determinism inappropriately.
In Section~\ref{sec:motivating-example}, we will give a real-world example of such bugs.

    \section{Motivating Example}\label{sec:motivating-example}

In this section, we present a code snippet adapted from Giveth\cite{giveth}, to illustrate the on-chain-off-chain synchronization in DApps, and an on-chain-off-chain synchronization bug~\cite{giveth_bug}.

\begin{lstlisting}[language=JavaScript, caption={Withdraw Donation Functionality in the DApp Giveth}, float=*, label=giveth_example, escapechar=|]
function withdrawDonation() { // withdraw donation from one crowdfunding project
    contract.liquidPledging
        .withdraw(...arguments) // send withdraw transaction to smart contract |\label{giveth_example:line:3}|
        .once('transactionHash', hash => { // function called when transaction is sent |\label{giveth_example:line:4}|
            let txHash = hash; // |\label{giveth_example:line:5}|
            updateExistingDonation(existingDonation, amount); // update existing donation after withdraw |\label{giveth_example:line:6}|

            const withdrawRecord = {...}; // create a new withdraw record |\label{giveth_example:line:8}|
            feathersClient
                .service('/donations')
                .create(withdrawRecord) // save the withdraw record in the centralized database |\label{giveth_example:line:11}|
                .catch(onError);
        });
}
\end{lstlisting}


Listing~\ref{giveth_example} shows a code snippet for the ``withdraw donation'' functionality in the front-end client of Giveth, which allows users to withdraw the donations they receive.
The front-end client of Giveth uses web3.js~\cite{web3}, an official Ethereum JavaScript library, to send a transaction to an underlying smart contract ``liquidPledging'' and call its ``withdraw'' function, as seen in line~\ref{giveth_example:line:3}.
The client registers a callback for handling the ``transactionHash'' event, which occurs when the transaction is added to the blockchain transaction pool (line~\ref{giveth_example:line:4}). 
Since the transaction is to modify the on-chain state concerning the donation amount, the off-chain state is also updated accordingly (line~\ref{giveth_example:line:6}).
After the update, a corresponding withdraw record is created and stored in an off-chain centralized database (line~\ref{giveth_example:line:11}).

The code in Listing~\ref{giveth_example} may not work correctly because the ``withdraw donation'' transaction may remain in the transaction pool indefinitely or be dropped without execution. 
The occurrence of the ``transactionHash'' event only signifies that a transaction has been added to the transaction pool rather than the execution of the transaction.
Even if the transaction has been executed, it can still be reversed later.
Therefore, it is possible that the off-chain state is updated while the on-chain state remains unchanged (i.e., when the withdraw transaction is dropped or reversed).
Such state inconsistencies can lead to many undesirable consequences.
For example, let us consider the following scenario.
A user of Giveth sends a ``withdraw donation'' transaction to the smart contract, but the transaction is dropped by the blockchain.
However, due to the bug, the Giveth client incorrectly shows that the donation has been withdrawn.
The user may decide to donate the withdrawn cryptocurrency to another community.
In such a scenario, the new donation transaction may fail, causing unnecessary loss of transaction fees and poor user experience. 

It is difficult to expose on-chain-off-chain synchronization bugs using existing DApp testing tools.
For example, Ganache~\cite{consensyssoftwareinc.Ganache2021} is one of the most commonly used blockchain environments for testing Ethereum-based DApps.
Like testing on centralized databases, transactions sent to Ganache blockchain are executed immediately.
Therefore, when testing Giveth using Ganache, the on-chain and off-chain states are always consistent.
The bug mentioned above can never be detected. 

To tackle this problem, we propose to model the non-determinism in transaction execution using the transaction lifecycle.
Our approach simulates the non-deterministic transaction execution process and drives DApps to traverse each transaction's lifecycle systematically.
We also propose effective test oracles to detect inconsistencies between the on-chain and off-chain states automatically.

    \section{Methodology}\label{sec:methodology}

This section first proposes a state transition model to capture the lifecycle of transactions on the Ethereum blockchain.
We then discuss the challenges in on-chain-off-chain synchronization and identify two types of synchronization bugs that may arise in DApps.
After that, we propose our framework, \darcher, to detect \bugs in DApps.

\subsection{Transaction Lifecycle}\label{subsec:transaction-lifecycle}
\begin{figure}[ht]
    \centering
    \includegraphics[scale=0.85]{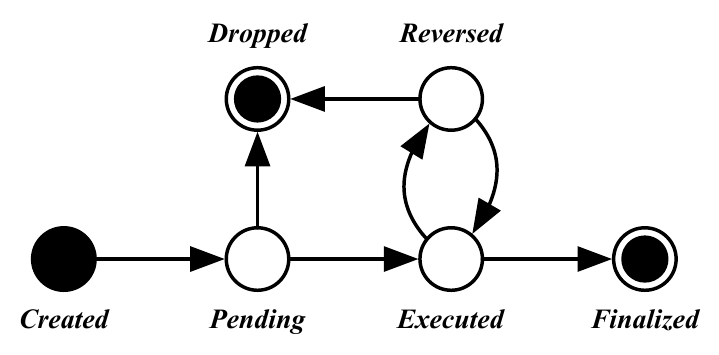}
    \caption{State Transition Model of the Ethereum Transaction Lifecycle 
    }
    \label{fig:transaction_lifecycle_model}
\end{figure}

To detect on-chain-off-chain synchronization bugs, we model the non-determinism in the lifecycle of a transaction on the Ethereum blockchain using a state transition model as shown in Fig.~\ref{fig:transaction_lifecycle_model}.
A transaction starts its lifecycle at the \CreatedState state by constructing the required arguments at the DApp's off-chain layer. 
After that, the transaction is sent to the blockchain and transits to the \PendingState state, meaning that the transaction is added to the transaction pool awaiting execution on the blockchain.
A transaction may remain in the \PendingState state indefinitely as miners are free to select more profitable transactions for execution.


A \PendingState transaction can be dropped and transit to the \DroppedState state in two cases.
First, users may send a duplicated transaction to override the previous transaction or offer a higher fee to increase the chance of execution~\cite{markouEIP2831TransactionReplacement2021}.
Second, the transaction may be deleted silently by miners due to the capacity limit of miners' transaction pool or malicious behaviors of miners.
In the latter case, the DApp is not informed that a transaction has been dropped.
It is also hard for the DApp to proactively check whether a transaction is dropped on the blockchain or not.
A  \PendingState transaction transits to the \ExecutedState state when it is executed and included in a block. 
As discussed in Section~\ref{sec:background}, an executed transaction can be reversed when the blockchain is reorganized.
When it happens, the transaction will transit from the \ExecutedState state to the \ReversedState state and be put back to the transaction pool, awaiting execution. 
Similar to the \PendingState state, transactions in the \ReversedState state can also be dropped.\footnote{
\addition{
We distinguish the \PendingState and \ReversedState states because they are separately handled by the Ethereum official library, web3.js~\cite{web3} under different JavaScript events: \textit{transactionHash} and \textit{changed}. 
}
}


While in theory executed transactions can be reversed,
in practice, a transaction whose execution has been logged by a block with a sufficient number of confirmations can be considered finalized, i.e., the transaction will transit from the \ExecutedState state to the \ConfirmedState state.
The number of confirmations required varies according to the security requirements of the DApp~\cite{how_many_confirmations}.
To avoid problems caused by reversed transactions, 
one common practice adopted by DApp developers is to use the result of a transaction at the off-chain layer only after the transaction transits to the \ConfirmedState state~\cite{gervaisSecurityPerformanceProof2016}.

To estimate the frequency of transaction state transitions, we collect Ethereum traffic data from Etherscan~\cite{etherscan} and an Ethereum full node maintained by us.
We find that the average number of transactions submitted to Ethereum per second is \txSubmitRate. 
The average number of transactions executed by Ethereum per second (TPS) is \TPS, which indicates that around half of the transactions cannot be immediately executed after submission.
We also monitor the Ethereum mainnet for over 4 months and observe that chain reorganization happens every \uncleHeightInterval blocks, i.e., every \uncleTimeInterval minutes.\footnote{\addition{
The frequency of chain reorganizations is calculated by the total number of canonical blocks mined (or the total time elapsed) divided by the total number of invalidated blocks within the period that we monitor the Ethereum mainnet. 
}}
The average number of transactions reversed per hour due to such chain reorganizations is \numRevokedTransactionsPerHour.
These statistics show that the drop and reverse rates of transactions on Ethereum are non-negligible, and it is necessary for DApps to consider the non-deterministic transaction execution to avoid on-chain-off-chain synchronization bugs.

\begin{figure*}[h!]
	\centering
	\subfigure[Type-\Rmnum{1}]{
		\label{fig:bug-diagram-1}
		\includegraphics[scale=0.9]{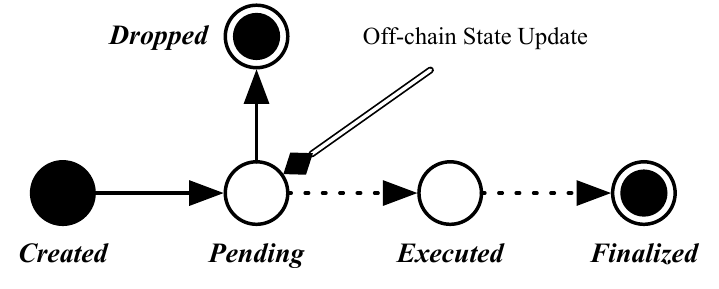}}
	\hspace{8mm}
	\subfigure[Type-\Rmnum{2}]{
		\label{fig:bug-diagram-2}
		\includegraphics[scale=0.9]{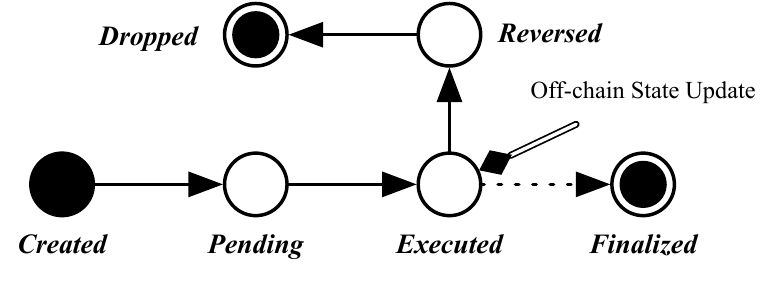}}
	\caption{Transaction Lifecycles for the Two Types of On-Chain-Off-Chain Synchronization Bugs. The solid arrows depict the actual state transitions, and the dashed arrows depict the transitions assumed by DApps. 
	For Type-\Rmnum{1} bugs, the off-chain state is prematurely updated when the transaction is \PendingState. 
	For Type-\Rmnum{2} bugs, the off-chain state is updated when the transaction is \ExecutedState, but the updated state is not reverted when the transaction is reversed.
	}
	\label{fig:bug-diagram}
\end{figure*}

\subsection{On-Chain-Off-Chain Synchronization Bugs}
In Section~\ref{sec:motivating-example}, we have presented an on-chain-off-chain synchronization bug in a DApp named Giveth.
On-chain-off-chain synchronization bugs occur when DApps fail to maintain the consistency between their on-chain states and off-chain states.
In the Giveth bug, developers do not consider the situation in which a \PendingState transaction is \DroppedState by the blockchain, and inconsistencies are thus induced.
Such inconsistencies can result in erroneous off-chain states, which might mislead users.
Further operations based on these erroneous states can cause unexpected changes on the blockchain or waste transaction fees.
Since DApps are a new kind of software applications, the lack of understanding of state transitions in the transaction lifecycle may make it difficult for developers to assure proper on-chain-off-chain synchronization.
We observe that developers often assume that the transactions submitted by their DApp would be \ExecutedState and eventually \ConfirmedState on the blockchain while ignoring the situations where (1) a \PendingState or \ReversedState transaction is dropped silently by the blockchain, or (2) an \ExecutedState transaction is reversed due to chain reorganization.
As a result, on-chain-off-chain-synchronization bugs often arise.
In this paper, we focus on studying on-chain-off-chain synchronization bugs triggered in the two situations.
We refer to them as Type-\Rmnum{1} and Type-\Rmnum{2} bugs, respectively.
Fig.~\ref{fig:bug-diagram} illustrates how these two types of bugs could occur with our transaction lifecycle model.
We will introduce them in detail below.
\subsubsection{Type-\Rmnum{1} Bugs}
As Fig.~\ref{fig:bug-diagram-1} shows, Type-\Rmnum{1} bugs occur because the off-chain state is prematurely updated when the transaction is in \PendingState state, but in fact, the transaction is later dropped. 
The bug in Giveth as shown in Listing~\ref{giveth_example} is a Type-\Rmnum{1} bug.
Giveth immediately updates its off-chain state (lines 4--13) when the donation withdrawal transaction is submitted to the blockchain. 
It does not further check the state of the transaction and does not adequately deal with potential state changes.

\subsubsection{Type-\Rmnum{2} Bugs}
As Fig.~\ref{fig:bug-diagram-2} shows, Type-\Rmnum{2} bugs occur because the off-chain state is updated after the transaction transits to the \ExecutedState state, but the executed transaction is later reversed due to chain reorganization.
Issue \#8260 of Augur~\cite{augur8260}, a popular DApp for prediction markets~\cite{augur}, is a Type-\Rmnum{2} bug.
Augur updates its off-chain state when the transactions that create markets are executed but does not revert the off-chain state when the executed transactions are reversed.
Consequently, the front-end client of Augur would display markets that do not exist on the blockchain.
Any further operations on these markets will result in ``Market Not Found'' errors.

These two types of bugs are not well studied in the literature.
Existing techniques either focused on testing smart contracts~\cite{luuMakingSmartContracts2016,jiangContractFuzzerFuzzingSmart2018,grechMadMaxSurvivingOutofgas2018,wangDetectingNondeterministicPayment2019,nguyenSFuzzEfficientAdaptive2020}, which are only concerned with the on-chain layer of DApps, or do not fully consider the entire lifecycle of transactions when testing DApps~\cite{consensyssoftwareinc.Ganache2021}.
Take the popular DApp testing environment Ganache~\cite{consensyssoftwareinc.Ganache2021} as an example. When using Ganache, DApp developers are encouraged to configure the testing blockchain to execute transactions immediately~\cite{ganache_options}, while the possible state transitions from \PendingState to \DroppedState or from \ExecutedState to \ReversedState are ignored~\cite{ganache_limitation}.
As a result, Type-\Rmnum{1} and Type-\Rmnum{2} bugs can easily slip into real-world DApps.
This motivates us to propose \darcher, a testing framework to effectively detect the two types of on-chain-off-chain synchronization bugs in DApps.

\subsection{The \darcher Testing Framework}\label{subsec:testing-on-chain-off-chain-synchronization-bugs}

We can see from the above discussion that we need to systematically emulate transaction state transitions to trigger \bugs.
Guided by our transaction lifecycle model, \darcher controls the execution of each transaction in its testing environment to drive the transaction to traverse possible states.
However, triggering bugs alone is not sufficient.
Effective testing also requires oracles to judge the existence of bugs. 
In the following, we present our design of test oracles in \darcher and then explain how to leverage the oracles to detect \bugs during testing.

\subsubsection{Test Oracles}
\label{sec:oracles}
It is challenging to design oracles for detecting on-chain-off-chain synchronization bugs.
Although the on-chain and off-chain states in DApps need to be consistent, they are not necessarily identical.
Off-chain states are often maintained as a simplified version of the corresponding on-chain states to facilitate front-end user actions.
For instance, in our motivating example, Giveth's cache server processes on-chain donations and stores processed data instead of making an exact copy of the on-chain state.
The processed data contains additional information, such as the index of donations, to facilitate user queries.
However, such information is not saved on blockchain to reduce the cost of interacting with smart contracts.
In addition, updates of the off-chain state depend on both the changes to the on-chain state and the program logic of the off-chain layer.
Therefore, it is hard to specify the mapping between the on-chain and off-chain states 
as well as directly check whether the off-chain state is consistent with the on-chain state.
To address the challenge, we design oracles that check on-chain-off-chain state consistency by only comparing the off-chain states of the DApp when the concerned transaction is at different lifecycle stages. 
In this way, the mapping between on-chain and off-chain states is not required.
In the following, we present the test oracles.

\medskip
\noindent\textbf{Test oracle for Type-\Rmnum{1} bugs.}
For Type-\Rmnum{1} bugs, we observe that, when a transaction $t$ is added to the transaction pool (i.e.,\ transits from the \CreatedState state to the \PendingState state), the DApp should not prematurely update the off-chain state as if $t$ is executed and finalized.
Since DApps are not informed when transactions transit from the \PendingState state to the \DroppedState state, any premature updates of the off-chain state are likely to remain when the transactions are dropped on the blockchain.
Assertion~1 helps detect such bugs.


\begin{assertion}
\label{assertion1}
For each transaction $t$,
    $\sigma(t,\CreatedState) \neq \sigma(t,\ConfirmedState)$ implies $\sigma(t,\PendingState) \neq \sigma(t,\ConfirmedState)$.
\end{assertion}

In the above formulation, $\sigma(t,s)$ denotes the off-chain state of the DApp under test when the transaction $t$ is at the transaction lifecycle state $s$.
The clause $\sigma(t,\CreatedState) \neq \sigma(t,\ConfirmedState)$ means that the transaction $t$ results in an update to the off-chain state.
The clause $\sigma(t,\PendingState) \neq \sigma(t,\ConfirmedState)$ specifies that the off-chain state should not be updated as if the transaction has been \ConfirmedState when the transaction has been just sent to the transaction pool.
Note that this assertion allows DApps to make changes to the off-chain state when the concerned transaction is \PendingState, but the changes should not indicate that the transaction has been \ConfirmedState.
Violations of Assertion~\ref{assertion1} indicate the existence of Type-\Rmnum{1} bugs.

\medskip
\noindent\textbf{Test oracle for Type-\Rmnum{2} bugs.}
When a transaction is executed (i.e., transits from the \PendingState state to the \ExecutedState state), some DApps may update their off-chain states.
Type-\Rmnum{2} bugs occur when the executed transaction is reversed due to chain reorganization, but the updated off-chain state is not reverted accordingly.
\darcher leverages Assertion~2 to check for Type-\Rmnum{2} bugs.

\begin{assertion}
\label{assertion2}
For each transaction $t$, $\sigma(t,\PendingState) = \sigma(t,\ReversedState)$.
\end{assertion}

Since reversed transactions are put back to the transaction pool, the off-chain state of the DApp when transaction $t$ is at the \PendingState state should be the same as that when $t$ is at the \ReversedState state.
Violations of Assertion~\ref{assertion2} indicate the existence of Type-\Rmnum{2} bugs.

\subsubsection{Lifecycle Emulation \& Assertion Checking}
\label{subsec:detecting-on-chain-off-chain-synchronization-bugs}
As discussed earlier, in existing blockchain testing environments such as Ganache~\cite{consensyssoftwareinc.Ganache2021}, transactions are directly \ExecutedState and \ConfirmedState once submitted to the blockchain.
These transactions are never dropped or reversed, and thus Type-\Rmnum{1} and Type-\Rmnum{2} bugs cannot be triggered.
To address the limitation, \darcher implements a blockchain environment that can control the state of transactions.
Instead of executing a transaction immediately after it is submitted to the blockchain, \darcher drives the transaction to traverse lifecycle states in the following order:
$\CreatedState \rightarrow \PendingState \rightarrow \ExecutedState \rightarrow \ReversedState \rightarrow \ExecutedState \rightarrow \ConfirmedState$.
Such a traversal allows our proposed oracles to be evaluated for each transaction to detect the two types of on-chain-off-chain synchronization bugs.
Specifically, when there is a state transition, \darcher fetches and stores the off-chain state of the DApp under testing.
After the state traversal terminates, the fetched off-chain state is checked against the Assertions~\ref{assertion1} and~\ref{assertion2}.
If there is any assertion violation, \darcher will report a bug.
More details of bug detection will be further introduced in Section~\ref{sec:darcher}.

    \section{Implementation}\label{sec:darcher}

\begin{figure}[ht]
    \centering
    \includegraphics[scale=0.5]{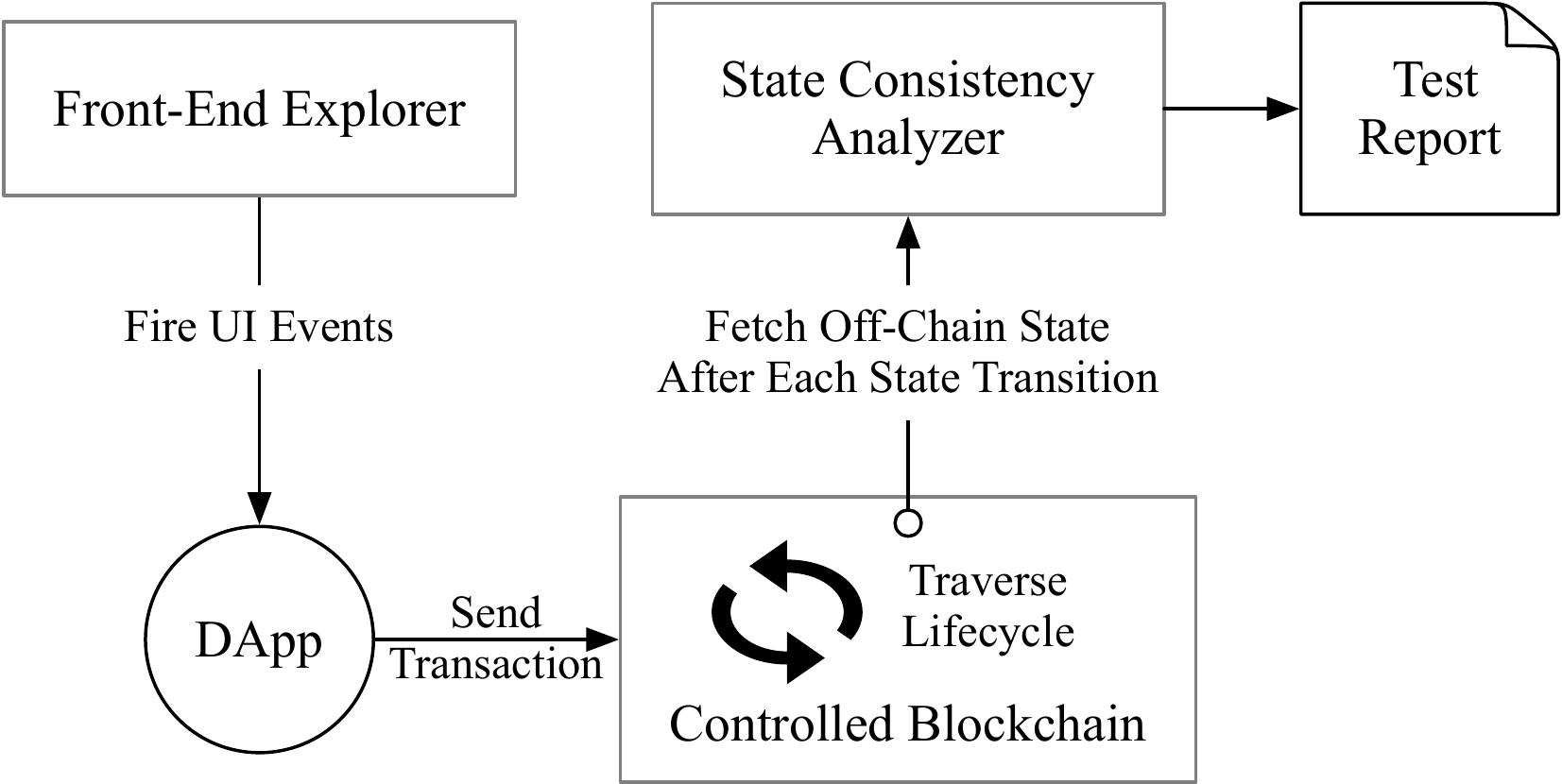}
    \caption{Overview and Workflow of \darcher}
    \label{fig:overview}
\end{figure}

Fig.~\ref{fig:overview} shows an overview of \darcher.
Given a DApp, the front-end explorer fires UI events to exercise the DApp to explore the functionalities that involve sending transactions.
Once a transaction is sent, \darcher leverages a controlled blockchain to execute it and traverse its lifecycle states in the order mentioned in Section~\ref{subsec:detecting-on-chain-off-chain-synchronization-bugs}.
During the state traversal process, \darcher keeps fetching the off-chain state of the DApp, whenever there is a state change. 
The state consistency analyzer then checks such collected off-chain states to detect bugs. 

We have open-sourced \darcher on GitHub\footnote{\url{https://github.com/Troublor/darcher}}.
In the following, we present more details on how \darcher is implemented. 

\subsection{The Front-End Explorer}
As mentioned earlier, the front-end explorer fires UI events to exercise DApps.
In practice, developers can use any applicable tool or manually write test cases for purpose.
In the current implementation of \darcher, we choose to integrate a popular web testing tool, Crawljax~\cite{mesbahInvariantBasedAutomaticTesting2012}, to generate GUI events to test DApps, which are often web-based applications.
Crawljax infers a state-flow graph when testing a web application.
GUI events are fired at the states that can transit to unvisited states. 
The state-flow graph is updated whenever new states are discovered during testing.
The exploration will stop when all of the states in the graph have been visited.
Such model-based testing can help exercise DApps to interact with the blockchain.

\subsection{The Controlled Blockchain}


We implement a controlled blockchain in \darcher based on Geth~\cite{geth}, a popular Ethereum client.
When the DApp under test submits a transaction to our controlled blockchain, \darcher will drive the transaction to traverse its lifecycle states according to the order mentioned in Section~\ref{subsec:detecting-on-chain-off-chain-synchronization-bugs}.

\subsection{The State Consistency Analyzer}
\label{sec: state consistency analyzer}
\aaron{The reviewer A stated that ``I would have liked to see more of the specifics of 5.3. It's not like a simple FSM has greater conceptual novelty than an elegant configuration language.''. However, I don't quite understand what he/she meant here. The sentence seems broken.}
As mentioned in the overview, during testing, \darcher keeps collecting the off-chain state data for bug detection.
Different DApps may maintain the off-chain states in different ways, e.g.,~using various databases such as MySQL and MongoDB, or browser storage such as IndexedDB and localStorage.
It is hard to design a tool to automatically identify runtime values representing the off-chain states for all \DApps.
As a workaround, we build our off-chain state fetcher for different data storage, including databases, browser storage, and HTML elements.
Users of our tool only need to configure a few rules to specify the variables and fields that constitute the off-chain state of their DApps \addition{(e.g., specifying column names in databases or regular expressions to include or exclude table columns)}.

To minimize the runtime overhead, \darcher does not instrument DApps. 
Since it is hard to determine when a DApp finishes updating its off-chain state at runtime, \darcher would wait for a period of time when a transaction's lifecycle state changes before fetching the off-chain state. 
The waiting time is configurable in \darcher. 

After the state traversal process completes, \darcher checks the fetched off-chain states to identify the two types of on-chain-off-chain synchronization bugs according to the two oracles proposed in Section~\ref{sec:oracles}.


    \section{Evaluation}
\label{sec:evaluation}

We evaluate \darcher on real-world web-based \DApps. Specifically, we investigate the following research questions:

\begin{itemize}[leftmargin=*, itemsep = 3pt, topsep = 3pt]
    \item \textbf{\namedlabel{RQ:effectiveness}{RQ1} (Bug Detection Capability)}: Can \darcher effectively detect \bugs in real \DApps?
    \item \textbf{\namedlabel{RQ:oracle}{RQ2} (Efficacy of Our Oracles)}: How effective are our proposed oracles? Can \bugs be detected using existing oracles?
    \item \textbf{\namedlabel{RQ:usefulness}{RQ3} (Usefulness)}: Can \darcher detect \bugs that are useful to developers?
\end{itemize}

\subsection{Subjects}\label{subsec:subjects}
To collect subjects for evaluation, we search for Ethereum \DApps on GitHub using the keyword \textit{ethereum} with constraints \textit{stars:>=100, language:JavaScript} or \textit{language:TypeScript}. 
The first constraint is to ensure the popularity of the selected subjects.
The second constraint is to help find web-based DApps.
The search returns 254 projects.
We manually check each project to exclude those that are not web applications or those that are libraries, blockchain clients, block explorers, transaction trackers, frameworks, operating systems, and so on.
Projects that are tagged as deprecated or archived are also excluded because they are not actively maintained.
Furthermore, we exclude those that are claimed to be examples, tutorials, or starters, as we are interested in real DApps rather than toy examples.
After this filtering process, 21 DApps remain.
Finally, \darcher is a dynamic testing framework, which requires executing transactions of smart contracts on a local controlled blockchain.
\addition{
Although a number of DApps provide the source code of smart contracts and web applications, many of them lack instructions to deploy their smart contracts on a local blockchain.
}\footnote{
    \addition{Deploying contracts not only requires sending contract creation transactions but also involves other specific transaction to initialize contract states, e.g.,~linking to other existing contracts, setting configurations, etc.}
}
\addition{
They expect users to deploy the open-source web applications on Ethereum public testnets or mainnet, where contracts have been well-deployed.
}\footnote{
    \addition{As an example, the web interface of a famous decentralized exchange DApp, Uniswap, only works on testnets and will not work on other blockchains, as stated by developers~\cite{uniswapUniswapinterface2021}. }
}
\addition{Therefore,} we exclude those projects that we fail to deploy on our local blockchain by following the provided instructions.
After the above filtering, we collect a total of \numSubjects popular real-world \DApps for experiments.
This is in line with the finding in an empirical study made by Wu et al.~\cite{wuFirstLookBlockchainbased2019} that few \DApps are fully open-source.
Table~\ref{tab:subjects} provides the information of these subjects.
The selected subjects differ in scale. 
The smallest DApp has less than 1,000 lines of code, while the largest DApp has more than 30,000 lines of code.
The purposes of the subjects are also quite diversified.

\subsection{Experiment Design}\label{subsec:experiment-design}

In this subsection, we explain how we set up our experiments, derive off-chain states, and establish the ground truth to validate bug detection results. 
We also introduce the baselines, against which \darcher will be compared. 

\textbf{Experiment Setup.}
We deploy the subjects locally on our controlled blockchain. 
We set the \darcher's waiting time for DApps to update their off-chain states to be 15 seconds, which is aligned with the average block interval time on Ethereum mainnet~\cite{etherscan}.
We run \darcher on each \DApp for one hour.
We observe that this is sufficient for \darcher to reach the saturation of test coverage during experiments. 
We repeat the testing process ten times for each \DApp to mitigate the randomness of Cralwjax.



\textbf{Off-Chain State Derivation.} One challenge in the experiments is to identify the data fields that compose off-chain states.
As we have discussed in Section~\ref{sec: state consistency analyzer}, different DApps may maintain off-chain states in different ways.
\addition{
Essentially, \darcher requires developers to manually specify which data fields or variables in the \DApp constitute the off-chain state.
However, such specification is unavailable for our evaluation.
}
As a workaround, we manually explore each DApp and derive the off-chain state by including those data fields that are updated when transactions are directly executed (i.e., going through the states $\CreatedState \rightarrow \PendingState \rightarrow \ExecutedState \rightarrow \ConfirmedState$).
The intuition behind is that we assume developers have tested their DApp to assure that the off-chain state is properly updated when transactions undergo such ``normal'' executions.
\addition{Note that \darcher does not integrate this off-chain state derivation mechanism in that the aforementioned assumption does not necessarily hold for all \DApps.}

\textbf{Result Validation.} 
To evaluate the precision and recall of \darcher for each DApp, we manually reproduce all transactions generated during testing and make them go through the lifecycle to identify potential inconsistencies between the on-chain and off-chain states.
This process is independently performed by two authors.
The results are cross-checked for consistency.
It is worth mentioning that the recall metric we used here is for evaluating \darcher's capability of catching Type-\Rmnum{1} and Type-\Rmnum{2} bugs once they arise during the processing of transactions.
\aaron{The reviewers complain that the ``recall'' is not true recall (since we don't have ground truth). Do we need to explain here that the recall is only an approximation?} \scc{We can clarify it.}
We do not aim to evaluate how many of all possible \bugs can be detected by \darcher since it is hard to obtain the ground truth. 
Not only that, the Crawljax front-end explorer may not be able to trigger all possible transactions during testing.

\textbf{Baselines.} 
Since there is no prior work for detecting \bugs in \DApps, we construct two baseline methods by replacing oracles used by \darcher with the ones used for detecting smart contract vulnerabilities~\cite{jiangContractFuzzerFuzzingSmart2018} and web application faults~\cite{biagiolaDiversitybasedWebTest2019}.
Specifically, Baseline-\Rmnum{1} would report bugs if the execution of a transaction violates the assertions defined by ContractFuzzer~\cite{jiangContractFuzzerFuzzingSmart2018}.
\aaron{Reviewers complains that the Baseline-\Rmnum{1} is a complete strawman. Do we consider to discard Baseline-\Rmnum{1}?} \scc{I think we can keep it.}
Baseline-\Rmnum{2} would report bugs if runtime errors occur in the JavaScript console during the testing process~\cite{biagiolaDiversitybasedWebTest2019}.
Except for the differences in oracles, the two baseline methods use exactly the same tests for each \DApp and traverse the lifecycle of each transaction in the same way as \darcher.
We do not compare \darcher with existing blockchain testing environments such as Ganache~\cite{consensyssoftwareinc.Ganache2021}, since none of them is able to emulate the lifecycle of transactions, and thus neither Type-\Rmnum{1} nor \mbox{Type-\Rmnum{2}} bugs can be triggered by them.
\begin{table}[]
    \centering
    \caption{The DApps Used in Our Experiments}
    \label{tab:subjects}
    \resizebox{\linewidth}{!}{%
    \begin{tabular}{@{}lrrrl@{}}
        \toprule
        DApp                                    & Stars & LOC (JS) & Commits & Purpose\\ \cmidrule(){1-1} \cmidrule(l){2-5}
        AgroChain~\cite{AgroChain}                         & 106            & 291               & 40               & Agricultural Supply Chain\\
        Augur~\cite{augur}                                 & 264            & 68,972            & 40,971            & Prediction Markets\\
        DemocracyEarth~\cite{DemocracyEarth}               & 1,305           & 29,149            & 3,394             & Governance for DAOs\\
        ETH Hot Wallet~\cite{eth-hot-wallet}               & 197            & 7,295             & 284              & Wallet\\
        Ethereum Voting Dapp~\cite{ethereum_voting_dapp} & 348            & 238               & 31               & Voting\\
        Giveth~\cite{giveth}                               & 257            & 30,708            & 3,131             & Charitable Donation\\
        Heiswap~\cite{heiswap}                             & 100            & 1,687             & 87               & Anonymous Transfer\\
        MetaMask~\cite{metamask}                           & 4,091           & 81,898            & 11,072            & Wallet\\
        Multisender~\cite{multisender}                     & 165            & 1,421             & 33               & One-to-Many Transfer\\
        PublicVotes~\cite{publicvotes}                     & 149            & 628               & 34               & Voting\\
        TodoList Dapp~\cite{ethereum-todolist}         & 127            & 1,076             & 28               & Todo List\\ \bottomrule
    \end{tabular}%
    }
\end{table}

\subsection{RQ1: Bug Detection Capability}\label{subsec:results-for-rq:effectiveness}

Table~\ref{tab:detection-results} presents the results of \darcher and baselines, including the overall test coverage, precision, recall, and accuracy. 
Since \darcher aims to test the correctness of synchronization between the on-chain and off-chain layers in DApps during the transaction execution process, we measure the test coverage in terms of how well transaction submission API call sites are covered during testing\footnote{We are able to measure the coverage of transaction submission API call sites because DApps use designated APIs~\cite{web3-sendTransaction, web3-sendSignedTransaction, web3-contractSendTransaction} provided by the official libraries (e.g.,\ web3.js~\cite{web3}) of Ethereum to interact with the blockchain. 
}.
Such API call sites are essentially the ``starting point'' of on-chain-off-chain synchronization. 
Their coverage could indicate the diversity of the tests generated by the Front-End Explorer.
Note that since some functionalities relying on Ethereum public blockchain services (e.g.,\ Uniswap~\cite{uniswap}) are unavailable on the controlled blockchain of \darcher, API call sites for such functionalities are not considered in our coverage measurement.
In the following, we will discuss the experiment results in detail.

\textbf{Coverage.} \darcher triggers a total of 3,134 transactions and covers 70\% of the transaction submission API call sites when testing the 11 DApps.
It achieves an overall accuracy of \accuracy in deciding whether a transaction handling process contains bugs or not.  

\textbf{False Positives.}
Altogether, there are 385 transactions violating Assertion~\ref{assertion1} and 1,862 transactions violating Assertion~\ref{assertion2}.
There is no transaction violating both assertions.
Among these transactions, there are 16 exhibiting Type-\Rmnum{1} bugs, which are found to be false positives (FPs).
There are no false positives for those detected to exhibit Type-\Rmnum{2} bugs. 
We find that all of the 16 FPs are related to the DApp Giveth.
They arise from the unexpected delays in updating the off-chain states in Giveth's cache server. 
For instance, when a transaction reaches the \PendingState state, Giveth makes a partial update to the off-chain state, with a flag indicating that the transaction is awaiting execution. 
The flag should be cleared when the transaction is \ConfirmedState.
However, it is not cleared within 15 seconds, so that \darcher fetches the same off-chain state as the one fetched when the transaction is \PendingState.
In this case, \darcher reports a violation of Assertion~\ref{assertion1}, but in fact, it is a false positive.
If \darcher waits for a longer period before fetching the off-chain state, the violation will not be reported.
Despite the FPs, \darcher still achieves an overall precision of \precision in detecting \bugs in the collected DApps.

\textbf{False Negatives.} There are 52 and 264 transactions exhibiting Type-\Rmnum{1} and Type-\Rmnum{2} bugs, respectively, which are missed by \darcher.
After investigating the corresponding transactions, we find two major reasons for the false negatives (FNs). 
First, \darcher may fetch the off-chain state before it is inappropriately updated, in the same way as mentioned in the previous Giveth example.
Second, transactions may depend on each other. 
An \bug may occur if a transaction $v$ is executed based on the interim result of another transaction $u$, while $u$ is reversed and dropped.
FN occurs if the DApp does not update the off-chain state for transaction $v$ (neither Assertion~\ref{assertion1} or \ref{assertion2} will be violated if the off-chain state is unchanged).
Despite such cases, the recall of \darcher is still quite high and reaches \recall in our experiments.



\begin{mdframed}[style=MyFrame]
    \textbf{Answer to~\ref{RQ:effectiveness}}:
    \darcher can effectively detect \bugs in \DApps with high precision (\precision), recall (\recall), and accuracy (\accuracy).
\end{mdframed}

\subsection{RQ2: Efficacy of Our Oracles}\label{subsec:results-for-rq:oracle}

We answer~\ref{RQ:oracle} by comparing \darcher with the two baseline methods that employ existing oracles.
The results of baselines are also presented in Table~\ref{tab:detection-results}.

As we can see from the table, Baseline-\Rmnum{1} only generates warnings for seven transactions.
After inspecting the seven transactions whose execution violates vulnerability assertions, we find that the underlying smart contract contains the exception disorder vulnerability according to the bug definition by Jiang et al.~\cite{jiangContractFuzzerFuzzingSmart2018}. 
However, these seven warnings are all FPs in terms of \bugs. 
Smart contract vulnerability oracles cannot detect any \bugs in the experiment.
This is because the detection of \bugs requires the examination of both on-chain and off-chain layers of a DApp, whereas the oracles for smart contract vulnerabilities examine only the on-chain layer.


Baseline-\Rmnum{2} is able to reveal some \bugs. 
For example, a bug~\cite{augur8260} in Augur results in a runtime error with message ``\textit{Uncaught (in promise) Error: execution reverted}'', when a transaction is reversed. 
However, the runtime error oracle is not effective compared to our proposed oracles.
Our experiments show that the overall precision, recall, and accuracy of Baseline-\Rmnum{2} are only 56.9\%, 10.1\% and 20.7\%, respectively, which are significantly worse than \darcher. 
In addition, the runtime error messages generated may not provide useful information about the root causes of \bugs.
For instance, the message ``Error: PollingBlockTracker - encountered error fetching block:'' generated in the testing of MetaMask gives little hint of the occurrence of a synchronization bug. 
\begin{mdframed}[style=MyFrame]
    \textbf{Answer to~\ref{RQ:oracle}}:
    Our proposed oracles significantly outperform the existing ones in terms of detecting \bugs.
\end{mdframed}

\begin{table*}[]
    \centering
    \caption{Experiment Results of \darcher and Baselines
   }
    \label{tab:detection-results}
    \resizebox{\textwidth}{!}{%
    \begin{threeparttable}
\begin{tabular}{@{}lrrrrrrrrrrrrrrrrrrrrrrr@{}}
\toprule
\multirow{3}{*}{\vspace{-5pt}DApp} & \multicolumn{1}{c}{\multirow{3}{*}{\begin{tabular}[c]{@{}c@{}}API\\ Call Site\\ Coverage\end{tabular}}} & \multicolumn{1}{c}{\multirow{3}{*}{\begin{tabular}[c]{@{}c@{}}Total\\ Txs.\end{tabular}}} & \multicolumn{9}{c}{\darcher}                                                                                                                     & \multicolumn{6}{c}{Baseline-\Rmnum{1}}  & \multicolumn{6}{c}{Baseline-\Rmnum{2}}                                            \\ \cmidrule(l){4-12} \cmidrule(l){13-18} \cmidrule(l){19-24}
\multicolumn{1}{c}{}                      & \multicolumn{1}{c}{}                                                                                    & \multicolumn{1}{c}{}                                                                      & \multicolumn{2}{c}{\hspace{10pt}TP} & \multicolumn{2}{c}{\hspace{6pt}FP} & \multicolumn{2}{c}{\hspace{6pt}FN} & \multirow{2}{*}{Pre.} & \multirow{2}{*}{Rec.} & \multirow{2}{*}{Acc.} & \multicolumn{6}{c}{Contract Vulnerability Oracle}   & \multicolumn{6}{c}{Runtime   Error Oracle}   \\
\multicolumn{1}{c}{}                      & \multicolumn{1}{c}{}                                                                                    & \multicolumn{1}{c}{}                                                                      & \Rmnum{1}  & \Rmnum{2} & \Rmnum{1}  & \Rmnum{2} & \Rmnum{1}  & \Rmnum{2} &                       &                       &                       & TP & FP & FN   & Pre. & Rec. & Acc. & TP  & FP  & FN   & Pre. & Rec. & Acc. \\ \cmidrule(){1-3} \cmidrule(l){4-12} \cmidrule(l){13-18} \cmidrule(l){19-24}
AgroChain                                 & 75.0\%                                                                                                  & 417                                                                                       & 60         & 160       & 0          & 0         & 40         & 0         & 100.0\%                     & 84.6\%                  & 90.4\%                   & 0  & 0  & 260  & -    & 0.0\%    & 37.6\% & 0   & 0   & 260  & -    & 0.0\%    & 37.6\% \\
Augur                                     & 66.7\%                                                                                                  & 164                                                                                       & 0          & 24        & 0          & 0         & 0          & 25        & 100.0\%                     & 49.0\%                  & 84.8\%                  & 0  & 7  & 49   & 0.0\%    & 0.0\%   & 65.9\% & 49  & 115 & 0    & 29.9\%  & 100.0\%    & 29.9\%  \\
DemocracyEarth                            & 100.0\%                                                                                                 & 78                                                                                        & 0          & 78        & 0          & 0         & 0          & 0         & 100.0\%                     & 100.0\%                     & 100.0\%                     & 0  & 0  & 78   & -    & 0.0\%    & 0.0\%    & 0   & 0   & 78   & -    & 0.0\%    & 0.0\%    \\
ETH Hot Wallet                            & 100.0\%                                                                                                 & 140                                                                                       & 140        & 0         & 0          & 0         & 0          & 0         & 100.0\%                   & 100.0\%                   & 100.0\%                   & 0  & 0  & 140  & -    & 0.0\%    & 0.0\%    & 58  & 0   & 82   & 100.0\%    & 41.4\% & 41.4\% \\
Ethereum Voting Dapp                    & 100.0\%                                                                                                 & 376                                                                                       & 0          & 140       & 0          & 0         & 0          & 236       & 100.0\%                     & 37.2\%                  & 37.2\%                  & 0  & 0  & 376  & -    & 0.0\%    & 0.0\%    & 0   & 0   & 376  & -    & 0.0\%   & 0.0\%   \\
Giveth                                    & 63.6\%                                                                                                  & 353                                                                                       & 27         & 0         & 16         & 0         & 11         & 0         & 62.8\%                  & 71.1\%                  & 92.4\%                  & 0  & 0  & 38   & -    & 0.0\%    & 89.2\% & 0   & 80  & 38   & 0.0\%    & 0.0\%    & 66.6\% \\
Heiswap                                   & 100.0\%                                                                                                 & 323                                                                                       & 0          & 322       & 0          & 0         & 0          & 1         & 100.0\%                  & 99.7\%                  &99.7\%                  & 0  & 0  & 323  & -    & 0.0\%    & 0.0\%    & 16  & 0   & 307  & 100.0\%    & 5.0\% & 5.0\% \\
MetaMask                                  & 100.0\%                                                                                                 & 225                                                                                       & 0          & 225       & 0          & 0         & 0          & 0         & 100.0\%                 & 100.0\%                 & 100.0\%                & 0  & 0  & 225  & -    & 0.0\% & 0.0\% & 1   & 0   & 224  & 100.0\%    & 0.4\%    & 0.4\%    \\
Multisender                               & 100.0\%                                                                                                 & 334                                                                                       & 0          & 332       & 0          & 0         & 0          & 2         & 100.0\%                     & 99.4\%                  & 99.4\%                  & 0  & 0  & 334  & -    & 0.0\%    & 0.0\%    & 0   & 0   & 334  & -    & 0.0\%    & 0.0\%    \\
PublicVotes                               & 100.0\%                                                                                                 & 289                                                                                       & 142        & 146       & 0          & 0         & 1          & 0         & 100.0\%                     & 99.7\%                     & 99.7\%                     & 0  & 0  & 289  & -    & 0.0\%   & 0.0\%    & 133 & 0   & 156  & 100.0\%    & 46.0\% & 46.0\% \\
TodoList Dapp                         & 100.0\%                                                                                                 & 435                                                                                       & 0          & 435       & 0          & 0         & 0          & 0         & 100.0\%                  & 100.0\%                  & 100.0\%                 & 0  & 0  & 435  & -    & 0.0\%  & 0.0\% & 0   & 0   & 435  & -    & 0.0\% & 0.0\% \\ \midrule
Total/Overall                                     & 70.0\%                                                                                                  & 3,134                                                                                      & 369        & 1,862      & 16         & 0         & 52         & 264       & 99.3\%                  & 87.6\%                  & 89.4\%                  & 0  & 7  & 2,547 & 0.0\%    & 0.0\%    & 18.5\% & 257 & 195 & 2,290 & 56.9\% & 10.1\%  & 20.7\% \\
Average & 77.3\% & 284.9 & 33.5 & 169.3 & 1.5 & 0 & 4.7 & 24.0 & 96.6\% & 85.5\% & 91.2\% & 0.0 & 0.6 & 231.5 & 0.0\% & 0.0\% & 17.5\% & 23.4 & 17.7 & 208.2 & 71.7\% & 17.5\% & 20.6\% \\
\bottomrule
\end{tabular}%
\begin{tablenotes}
    Txs. is short for transactions.
    We report the number of TPs, FPs and FNs for Type-\Rmnum{1} and Type-\Rmnum{2} bugs separately in the detection results of \darcher.
    We use Pre., Rec., and Acc. as abbreviations for precision, recall, and accuracy, respectively. Precision is marked as ``-'' when the tool reports no bugs. The average results are arithmetic means of the results for all subjects.
\end{tablenotes}
\end{threeparttable}
}
\end{table*}

\begin{table}[]
\centering
\caption{Real Bugs Detected by \darcher}
\label{tab:report-results}
\resizebox{\linewidth}{!}{%
\begin{threeparttable}
    \begin{tabular}{@{}lc|lc@{}}
        \toprule
        DApp & Issues                                                                          & \multicolumn{1}{l}{DApp} & Issues                     \\ \midrule
        AgroChain                & \#7, \#8                                                                        & Heiswap                  & \#29                       \\
        Augur                    & \#8260\textsuperscript{f}                                                       & MetaMask                 & \#10120\textsuperscript{c} \\
        DemocracyEarth           & \#561                                                                           & Multisender              & \#34\textsuperscript{c}    \\
        ETH-Hot-Wallet           & \#41                                                                            & PublicVotes              & \#11, \#12                 \\
        Ethereum Voting Dapp     & \#28                                                                            & TodoList Dapp            & \#5\textsuperscript{?}     \\
        Giveth                   & \#1103\textsuperscript{f}, \#1605\textsuperscript{f}, \#1792\textsuperscript{c} &                          &                            \\ \bottomrule
    \end{tabular}%
\begin{tablenotes}
            \item[f] The reported bugs have been confirmed and fixed.

            \item[c] The reported bugs have been confirmed.

            \item[?] The developers have asked for Pull Requests.

            \item Other issues have not received responses from developers.
        \end{tablenotes}
    \end{threeparttable}
}
\end{table}

\subsection{RQ3: Usefulness}\label{subsec:results-for-rq:usefulness}
We answer \ref{RQ:usefulness} by reporting bugs detected by \darcher to the developers and communicating with them. 
Although \darcher reports warnings for thousands of transactions in the experiments, lots of them are repeated explorations of the same functionalities in DApps.
To avoid overwhelming developers, we group the warnings with the same root cause into a single issue to report to the developers.
Table.~\ref{tab:report-results} lists the IDs of the GitHub issues, in which we report \bugs to the DApp developers.

In total, we have reported \numReportedBugs bugs, of which \numConfirmedBugs have been confirmed by developers, and \numFixedBugs have been fixed.
Developers provide positive feedback on our reported bugs.
For example, developers of Giveth respond ``Syncing two backend (cache server and blockchain) is a delicate and complex job, and we hope it is solved soon'', and the bugs in Giveth were fixed one and a half month after we reported them.
The comment indicates that the reported \bugs are real, and the on-chain/off-chain synchronization is complex.
Detection of the bugs is useful to developers in improving the quality of \DApps.

Due to the complexity of on-chain/off-chain synchronization, developers are likely to improperly handle scenarios where transactions are dropped or reversed. 
For instance, the developers of Giveth are aware of the possibility that transactions can be reversed after execution, and Giveth is designed only to update the off-chain state when transactions reach the \ConfirmedState state.
However, as discussed in Section~\ref{sec:motivating-example}, the handling of some transactions is still flawed, and the off-chain state is updated prematurely without waiting for the transaction to be finalized.
Another example is Augur, which maintains a rollback table that stores the metadata used to revert the off-chain state when a transaction is reversed on the blockchain.
However, the rollback table is cleared unexpectedly when the page is refreshed after the transaction is executed.
In such cases, the off-chain state does not get reverted when a transaction is reversed, causing inconsistencies between off-chain and on-chain states.
\darcher has been able to catch the above \bugs, and they have been confirmed and fixed by developers.
This demonstrates that \bugs could still occur in those DApps whose developers have already considered the non-determinism of transaction execution, and \darcher is able to help developers catch the hidden bugs.


\begin{mdframed}[style=MyFrame]
    \textbf{Answer to~\ref{RQ:usefulness}}:
    Among 15 bugs reported to developers, \numConfirmedBugs have been confirmed and \numFixedBugs have been fixed.
    Responses from developers show that \darcher is useful in detecting \bugs.
\end{mdframed}

    \section{Discussions}\label{sec:discussion}

\subsection{Synchronization Strategies}\label{subsec:strategies-of-on-chain-off-chain-synchronization}
\addition{
In the evaluation, we observe that some \DApps have considered the possibility that pending transactions can be silently dropped and that executed transactions can be reversed. \lili{reversed or reverted?}\aaron{We specially use ``reverse'' for transactions, as in the lifecycle model.}
However, \bugs are still detected in these \DApps.
This indicates that on-chain-off-chain synchronization is non-trivial and error-prone, highlighting the detection capability of \darcher.
}

To better understand how developers of \DApps synchronize on-chain and off-chain states, we further investigate the synchronization strategies adopted in the 11 \DApps used in our evaluation.
We observe three common strategies as follows.

\textbf{Periodic Polling.}
The most straightforward way to synchronize on-chain and off-chain states is to poll the on-chain state periodically.
For instance, DemocracyEarth~\cite{DemocracyEarth} registers a daemon task that periodically checks the on-chain state and updates the off-chain state accordingly.
This strategy effectively keeps the off-chain state consistent with the on-chain state during the lifecycle of transactions.
Nevertheless, periodic polling is inefficient if the \DApp is complicated.
If redundant synchronization work is performed repeatedly, a lot of communication and computation overheads will result.

\textbf{Passive Waiting.}
The passive waiting strategy copes with the non-determinism of transaction execution by updating the off-chain state only when a transaction reaches the \ConfirmedState state in its lifecycle.
That is to say, as long as the non-determinism still exists, i.e.,~transactions are not yet finalized, the \DApp does not update its off-chain state.
For instance, Giveth~\cite{giveth} adopts this strategy by counting the number of confirmations for each transaction after it is executed.
Only when a transaction has enough confirmations will Giveth update its off-chain state in the centralized cache server.
This strategy could save a lot of communication and computation overheads compared to the periodic polling strategy since the DApp only needs to track the transactions that are not yet finalized.
However, it could induce an inevitable delay in the \DApp, influencing user experiences because after the transaction is executed, users must wait until there are enough confirmations.

\textbf{Aggressive Updating.}
The aggressive updating strategy is another choice for \DApp developers.
This strategy is intended to keep the off-chain state closely synchronous with the on-chain state, which means the \DApp updates its off-chain state when transactions are executed, and reverts the off-chain state when transactions are reversed.
For instance, Augur adopts this strategy in its implementation.
When a transaction is sent to the blockchain, that is, in \PendingState state, the off-chain state is not updated. 
The update only takes place when the transaction is executed.
If the transaction is reversed due to blockchain reorganization, Augur will also revert its off-chain state accordingly.
This strategy offers better user experience than the passive waiting strategy.
Users can see the updates of the off-chain state immediately when their transactions are executed or reversed.
However, it is more error-prone to revert the off-chain state, which might involve many data fields, when transactions are reversed on the blockchain.

\subsection{Limitations and Future Work}
Our work is subject to two limitations. First, precise identification of the off-chain states is important to the test effectiveness of \darcher.
In our experiments, we assume that the off-chain states are appropriately updated when the transactions are completed in a straightforward manner.
This may not always hold.
Second, \darcher assumes that each update of an off-chain state, if it indeed happens, would be completed within a fixed time period, which is to be manually specified.
If the period is set to be too large, the time efficiency of \darcher is compromised.
If the period is set to be too small, \darcher may miss the detection of some \bugs. Furthermore, the period can vary across \DApps. 
A possible solution to these two limitations is to analyze the source code of DApps to determine what comprises the off-chain states and when they will be updated so that the efficiency and effectiveness of each transaction's analysis could be improved. 
However, the dynamic and reflective nature of JavaScript~\cite{staicuExtractingTaintSpecifications2020} imposes other challenges to perform a sound and complete analysis of DApps. 
As such, we leave these two limitations to be addressed in our future work. 

Furthermore, as discussed in Section~\ref{subsec:results-for-rq:effectiveness}, it could be the case that the consequence of \bugs is not reflected in DApps' off-chain states, for instance, sending a transaction dependent on the interim result of another transaction, which gets reversed or dropped.
Future work can be made to detect the \bugs exhibited in such scenarios. 

\subsection{Threats to Validity}\label{subsec:threats-to-validity}
The limited number of DApp subjects poses an external threat to the validity of our evaluation results.
We mitigate this threat by selecting popular real-world DApps of different sizes and purposes from GitHub to improve their representativeness. 
More subjects are needed to address this threat in future research fully.
\addition{
Leveraging Crawljax to exercise \DApps induces another external threat in that Crawljax is a randomized tool and may not trigger all possible transactions. 
We mitigate this threat by repeating the experiments ten times in order to increase the diversity of explored functionalities.
}

Threats to internal validity may arise from the way we interpret the experiment results. 
The specification of off-chain states for each DApp poses a threat, which we mitigate by mechanically deriving off-chain states with the assumption discussed in Section~\ref{subsec:experiment-design}. 
During the reproduction of transactions, we confirm that our assumption holds for all DApp subjects.
We also report our detected bugs to the DApp developers and ask for their feedback to confirm the effectiveness and usefulness of \darcher.


    \section{Related Work}
\label{sec:review}
This section briefly reviews the existing work related to the problem that \darcher aims to address.  

\subsection{DApp Development and Testing}
In 2017, Porru et al.~\cite{porruBlockchainOrientedSoftwareEngineering2017} introduced the concept of blockchain-oriented software engineering and pointed out the challenges and research directions on the development and testing. 
Since then, DApps, as a kind of blockchain-oriented software, started to attract attention from software engineering researchers.
Wessling et al.~\cite{wesslingHowMuchBlockchain2018,wesslingBlockchainTacticsBuilding2019} discussed the design choices of the architecture of a software that involves blockchain, showing the benefits and drawbacks of DApps.
Wu et al.~\cite{wuFirstLookBlockchainbased2019,wuEmpiricalStudyBlockchainbased2019} conducted empirical studies on Ethereum-based DApps to show the popularity, growth, development practice, cost, and the open-source status quo. 
They pointed out the research direction in the synchronization between the on-chain and off-chain layers of a DApp, but no further study has been conducted. 
Wu et al.~\cite{wuKayaTestingFramework2020} proposed a framework, Kaya, for testing DApps. 
However, their framework only provides tools to facilitate the manual creation and execution of test cases for DApp. 
Unlike our work, Kaya does not target any bugs specific to DApps and does not propose oracles to help automatically reveal bugs in DApps. 

\subsection{Smart Contract Testing}
Lots of studies have been conducted over the past several years to analyze and test smart contracts.
Various approaches have been proposed to detect vulnerabilities with symbolic execution~\cite{luuMakingSmartContracts2016,mossbergManticoreUserFriendlySymbolic2019}, fuzzing~\cite{jiangContractFuzzerFuzzingSmart2018,nguyenSFuzzEfficientAdaptive2020,kolluriExploitingLawsOrder2019,liuReGuardFindingReentrancy2018a,wustholzTargetedGreyboxFuzzing2020}, static analysis~\cite{wangDetectingNondeterministicPayment2019,grechMadMaxSurvivingOutofgas2018,brentVandalScalableSecurity2018,xueCrossContractStaticAnalysis2020,feistSlitherStaticAnalysis2019,tsankovSecurifyPracticalSecurity2018,muellerSmashingEthereumSmart2018}, mutation testing~\cite{chapmanDeviantMutationTesting2019}, or machine learning~\cite{liuSgramSemanticawareSecurity2018} techniques.
Multiple empirical studies have also been conducted to review and verify the effectiveness and efficiency of smart contract vulnerability analysis tools~\cite{ghalebHowEffectiveAre2020,durieuxEmpiricalReviewAutomated2020}.
However, these studies only focus on testing the on-chain layer of a DApp, that is, smart contracts. 
As shown in our evaluation, in which we adopt contract vulnerability oracles from ContructFuzzer~\cite{jiangContractFuzzerFuzzingSmart2018}, testing smart contracts only and neglecting the testing of the interaction between on-chain and off-chain layers cannot ensure the correctness of a DApp.
Our work, instead, takes the interaction between the on-chain and off-chain layers into consideration and can detect bugs during the synchronization of the two layers. 

\subsection{Test Case Generation for Web Applications}
Since \darcher targets web-based DApps, test case generation in web applications is relevant to our work. 
Mesbah et al.~\cite{mesbahInvariantBasedAutomaticTesting2012} proposed Crawljax, which can derive a state flow graph for web applications and generate tests to traverse the graph. 
Since Crawljax is well-maintained and capable of automatically generating test cases with good coverage to explore web applications, we integrate it into \darcher to trigger the interaction with smart contracts for testing DApps. 
Other test generation tools, such as \textsc{SubWeb}~\cite{biagiolaSearchBasedPath2017} and \textsc{DIG}~\cite{biagiolaDiversitybasedWebTest2019}, are usually built based on Crawljax, with the aim to further increase the coverage of generated tests. 
Since these tools require web applications to have page navigational models, which are not available in our DApp subjects, we did not integrate them into our framework.
Nevertheless, the DApp Front-End Explorer component of \darcher is loosely coupled with other components.
It is convenient to migrate to other web testing tools to explore functionalities of DApps and detect on-chain-off-chain synchronization bugs based on our proposed oracles. 


    \section{Conclusion}
\label{sec:conclusion}

In this paper, we study the development challenges of DApps caused by the non-deterministic transaction execution on the blockchain and the \bugs thus induced.
We propose \darcher, an automated testing framework to detect two common types of such bugs in DApps. 
Our experiments on 11 real-world DApps show that \darcher is effective in detecting \bugs and significantly outperforms the baseline methods in terms of precision, recall, and accuracy.
Feedbacks from real-world DApp developers also confirm the effectiveness and usefulness of \darcher.

%

\begin{acks}
    We sincerely thank Yuqing Xiao for the early-stage discussions towards this idea; Jialun Cao and Lu Liu for the discussions about the feasibility of our methodology; 
    Yongqiang Tian and Jiarong Wu for their efforts of proof-reading; and anonymous reviewers for their constructive comments.
    This work was supported by the National Natural Science Foundation of China (Grant No. 61932021), Hong Kong RGC/GRF (Grant No. 16207120) and Guangdong Provincial
    Key Laboratory (Grant No. 2020B121201001), China. 
    Lili Wei was supported by the Postdoctoral Fellowship Scheme of the Hong Kong Research Grant Council.
\end{acks}

    \bibliographystyle{ACM-Reference-Format}
    \balance
    \bibliography{bose,common}




%
%
%

\end{document}